# Electrical Contacts to Carbon Nanotubes Down to 1nm in Diameter

*Woong Kim, Ali Javey, Ryan Tu, Jien Cao, Qian Wang, and Hongjie Dai**

*Department of Chemistry and Laboratory for Advanced Materials, Stanford University,*

*Stanford, California 94305*

**Abstract**

Rhodium (Rh) is found similar to Palladium (Pd) in making near-ohmic electrical contacts to single-walled carbon nanotubes (SWNTs) with diameters $d > \sim 1.6$ nm. Non-negligible positive Schottky barriers (SBs) exist between Rh or Pd and semiconducting SWNTs (S-SWNTs) with $d < \sim 1.6$ nm. With Rh and Pd contacts, the characteristics of SWNT field-effect transistors and SB heights at the contacts are largely predictable based on the SWNT diameters, without random variations among devices. Surprisingly, electrical contacts to metallic SWNTs (M-SWNTs) also appear to be diameter dependent especially for small SWNTs. Ohmic contacts are difficult for M-SWNTs with diameters $\leq \sim 1.0$ nm possibly due to tunnel barriers.

---

* Email: hdai@stanford.edu

Contacts are an indispensable part of electrical devices of any electronic materials. Various types of metal contacts have been investigated for single walled carbon nanotubes, especially for semiconducting single-walled carbon nanotube field effect transistors (SWNT FETs) aimed at utilizing the advanced materials properties including high carrier mobility, ballistic transport and high compatibility with high-κ dielectrics.[1-8] Pd has been shown recently as an excellent contact material for both semiconducting[2] and metallic SWNTs[5] to facilitate the elucidation of intrinsic properties of nanotubes and optimization of SWNT FETs.

Here, we identify Rh as another metal capable of forming high-quality contacts to SWNTs. With Rh and Pd contacts, the main characteristics of a SWNT FET are largely predictable without random fluctuations between devices once the diameters of the nanotubes are known. This allows for a systematic investigation of the diameter dependence of Schottky barrier height at the contacts between Rh or Pd and SWNTs. Further, a systematic investigation is extended to metallic SWNTs with diameters down to $d \sim 1$nm. We observe diameter dependent contact barriers at the metal/M-SWNT contacts. The origin of this barrier and the need of ohmic contacts for very small diameter SWNTs are discussed.

Individual SWNTs ($d \sim 1$ to 3 nm) were synthesized by chemical vapor deposition[9] and integrated into back-gated (G) three terminal devices with Rh or Pd as source (S) and drain (D) contact electrodes.[10] The thickness of the gate dielectric $SiO_2$ layer ($t_{ox}$) was 10 nm for semiconducting SWNTs (as FETs) with a channel length (distance between S/D metal electrode edges) of $L = 200$ to 300 nm, unless specified otherwise. The diameters of nanotubes were measured by atomic force microscopy (AFM) topography and averaged along the tube length with an error of ~ 10 %. Devices with Rh and Pd contacts were annealed in argon (Ar) at 250 °C and 200 °C respectively and then passivated with poly(methyl methacrylate) (PMMA) for hysteresis elimination.[11] Metallic SWNT devices fabricated on $t_{ox}$=500 nm thick SiO$_2$/Si substrates were used for electrical transport characterization.

We found that similar to Pd, Rh can form transparent contacts to S-SWNTs with $d \geq 2$ nm and afford high on-state currents. Fig. 1a and 1b shows a Rh contacted FET of a $d \sim 2.1$ nm nanotube at room temperature with linear on-state resistance of $R_{on} \sim 30$ k$\Omega$, subthreshold swing of $S \sim 150$ mV/decade (Fig.1a), on-current of $I_{on} \sim 23$ µA (Fig. 1b), and $I_{on}/I_{min} \sim 10^5$ ($I_{min}$ is the minimum current point in $I_{ds}$ vs. $V_{gs}$ curve in Fig.1a). The high $I_{on} \sim 23$ µA is similar to the saturation current of same-length M-SWNTs[12] and indicative of near zero SB for p-channel hole transport across the contacts.[2]

While there is no appreciable positive SB to the valence band of $d > 2$ nm S-SWNTs, finite positive SBs exist and manifest in the device characteristics of smaller diameter SWNTs (with larger band gaps $E_g \sim 1/d$). For a typical S-SWNT with $d \sim 1.5$ nm, little ambipolar conduction is observed (for $t_{ox}$=10 nm) and the on-off ratio is very high $I_{on}/I_{off} \sim 10^7$ (Fig. 1c) owing to the larger band gap than 2 nm tube. The on-current of $I_{on} \sim 5$ µA for the 1.5 nm SWNT at $V_{ds} = -1$ V (Fig. 1d) is lower than $I_{on} \sim 20$ µA for larger tubes due to the existence of a positive SB.

We measured tens of SWNT FETs with Pd and Rh contacts and observed a systematic trend in on-currents vs. diameter for SWNTs in the range of $d \sim 1$ to 3 nm (Fig.2a), without random fluctuations between devices. The 'on-current' here-on is defined as the current $I_{ds}$ measured under a S/D bias of $V_{ds} = -1$ V and $V_{gs} - V_{th} = -3$ V ($V_{gs}$: gate voltage; $V_{th}$ threshold voltage). We found that the data points for Rh and Pd contacted devices happened to fall onto a similar trace (Fig.2a), indicating that Rh and Pd afford very similar contacts for SWNTs. For $d \geq 2$nm SWNTs, Pd and Rh contacts gave negligible SB to the p-channel and on-currents $\geq 20$ μA. For SWNTs with 1.6 nm<$d$<2nm, the on-current exhibited a gradual decrease with $d$. For d<~1.6nm nanotubes, a rapid current drop was observed for small diameter S-SWNTs (Fig.2a).

The lower on-currents for smaller diameter SWNTs are attributed to increase in Schottky barrier height to the valence band of nanotubes and the resulting current limitation by thermal activation over the SB.[10] We have attempted to estimate the p-channel SB height between Pd or Rh with S-SWNTs with a simple analytical expression $\Phi_{BP} \sim (\Phi_{CNT} + E_g/2) - \Phi_M$, where $\Phi_{CNT} \sim 4.7$ eV is work function of SWNTs,[13] $E_g \sim 1.1$ eV/$d$(nm) is the band gap of SWNTs,[14] and $\Phi_M \sim 5$ eV is the similar metal workfunction for Pd and Rh.[15] This classical formulism is used here since little Fermi level pinning exists at the metal-nanotube contact.[2] The estimated SB height is then

$$\Phi_{BP} = 0.56 \times \left(\frac{1}{d} - \frac{1}{2}\right) \text{ eV, for } d \leq 2 \text{ nm,} \tag{1}$$

with $\Phi_{BP}$=0 for a $d$~2nm SWNT and $\Phi_{BP}$~70meV for a $d$~1.6nm tube. Eq.(1) is satisfactory in interpreting our experimental result of $I_{on}$ vs. $d$ for S-SWNTs. The rapid decrease in $I_{on}$ for $d < 1.6$ nm S-SWNTs (Fig.2a) can be explained by the existence of a positive SB of $\Phi_{BP}$> ~70 meV according to eq.(1). That is, the SBs are $\Phi_{BP}$>~3$k_B$T (the degenerate limit) at room temperature for $d$<1.6nm S-SWNTs. This severely limits thermal activation and leads to low currents as observed experimentally for $d$<1.6 nm S-SWNTs (Fig.2a). Since the width of the SB is on the order of $t_{ox}$~10nm,[7] tunneling current through the barrier is also limited. Eq. (1) shall prove useful for estimating the SB heights between Rh or Pd and various diameter S-SWNTs and predicting device characteristics. Note that $\Phi_{BP}$ ~95 meV estimated using Eq. (1) for the $d$~1.5nm SWNT in Fig.1c is similar to that of $\Phi_{BP}$~90 meV estimated from its transfer characteristics using a method described by Appenzeller et al.[16]

Next, we turn to metallic SWNTs (with relatively weak gate dependence, Fig.3a) with Pd and Rh contacts. We find that the resistance of the M-SWNT devices is relatively insensitive to the length of the tubes in the range of $L$=100nm – 300 nm investigated, but sensitive to tube diameter especially for small nanotubes (Fig.4a). Large diameter M-SWNTs ($d \geq$1.5nm) can be well-contacted by Pd and Rh to afford two terminal resistance of $R$~20 k$\Omega$ at room temperature. Smaller $d$ M-SWNTs are more resistive (~ hundreds of kilo-ohms for $d$~1 nm tubes, Fig.4a) and exhibit lower currents down to 5μA (at $V_{ds}$=1V) compared to 20 μA for $d \geq$1.5nm tubes (Fig.4b). Non-linearity near zero-bias in the current-voltage characteristics of small M-SWNTs is noticed (Fig. 3b), a sign of non-ohmic contacts.

To investigate whether resistance of small tube devices is dominant by the contacts or diffusive transport due to defects along the tube length, we carried out measurements at low temperatures. For a typical $d$~1.2 nm M-SWNT, we observed clean and regular patterns of Coulomb oscillations at T=4 K (Fig.3c). The well defined Coulomb diamonds and appearance of discrete conductance-lines suggest the $L$~120 nm (inset of Figure 3a) tube as a single coherent quantum-wire without significant defects breaking the nanotubes into segments. The charging energy of the nanotube is $U$ ~ 50 meV with a discrete level spacing of ~13 meV due to quantum confinement along the length.[17] The resistance of the system is dominant by contacts rather than defects in the SWNT, indicating significant barriers at the contacts of metal/small $d$ M-SWNT.

The $I_{on}$ vs. $d$ curves for M- and S-SWNTs are offset along the diameter-axis (Fig.4b) with low currents in S-SWNTs and M-SWNTs for $d$<~1.6 nm and $d$< ~1.0 nm tubes respectively. The existence of SBs for small S-SWNTs is responsible for lower currents than in same-diameter M-SWNTs. However, origin of the observed contact barrier for small M-SWNTs is not well understood. It is possible that a tunneling barrier (independent of SB in the case for S-SWNTs) exists between metal atoms and nanotube arising from the metal-SWNT chemical bonding configuration.[18] This tunneling barrier may depend on metal type (seen theoretically)[18] and also SWNT diameter. Our M-SWNT data shows larger height or width of the tunnel barrier at the metal-tube contacts for smaller tubes (Fig. 4). This causes serious contact resistance dominance for $d$<~1.2nm M-SWNTs.  For S-SWNTs with decreasing diameter (Fig. 4b), a positive SB develops prior to the manifestation of tunnel barrier and is the dominant source of contact resistance for $d$<2nm S-SWNTs. Chemical doping can be invoked to suppress the positive SBs as shown recently.[8]  However, tunnel barrier appears to be independent of doping and causes non-ohmic contact for $d$<1.2nm S-SWNTs even under chemical doping, as found in our doping experiments (A. Javey, H. Dai, unpublished results).

In summary, diameter dependent contact phenomena are observed for metallic and semiconducting nanotubes related to Schottky and tunnel barriers. Ohmic contacts to very small ($d$≤1nm) nanotubes is currently a key challenge and requires developing strategy to eliminate the large tunnel barriers to small SWNTs. Various chemical synthesis methods are known to produce very small tubes predominantly in the 0.7-1.2 nm range.[19] An ohmic contact solution for down to 0.7nm tubes will be needed in order to enable high performance electronics with these materials. To achieve this goal, systematic experiments and theoretical understanding will be needed including continued search for an optimum contact metal.

We acknowledge insightful discussions with K. J. Cho. This work was supported by MARCO MSD Focus Center.

8**REFERENCES**

1. A. Javey, H. Kim, M. Brink, Q. Wang, A. Ural, J. Guo, P. McIntyre, P. McEuen, M. Lundstrom, and H. Dai, Nature Materials **1,** 241 - 246 (2002).
2. A. Javey, J. Guo, Q. Wang, M. Lundstrom, and H. J. Dai, Nature **424,** 654-657 (2003).
3. S. Rosenblatt, Y. Yaish, J. Park, J. Gore, V. Sazonova, and P. L. McEuen, Nano Lett. **2,** 869-915 (2002).
4. S. Wind, J. Appenzeller, R. Martel, V. Derycke, and P. A. P, Appl. Phys. Lett. **80,** 3817-3819 (2002).
5. D. Mann, A. Javey, J. Kong, Q. Wang, and H. Dai, Nano Lett. **3,** 1541 - 1544 (2003).
6. T. Durkop, S. A. Getty, E. Cobas, and M. S. Fuhrer, Nano Lett. **4,** 35-9 (2004).
7. A. Javey, J. Guo, D. B. Farmer, Q. Wang, D. W. Wang, R. G. Gordon, M. Lundstrom, and H. J. Dai, Nano Lett. **4,** 447-450 (2004).
8. A. Javey, R. Tu, D. B. Farmer, J. Guo, R. G. Gordon, and H. J. Dai, Nano Lett. **5,** 345-348 (2005).
9. J. Kong, H. Soh, A. Cassell, C. F. Quate, and H. Dai, Nature **395,** 878 (1998).
10. C. Zhou, J. Kong, and H. Dai, Appl. Phys. Lett. **76,** 1597 (1999).
11. W. Kim, A. Javey, O. Vermesh, Q. Wang, Y. Li, and H. Dai*, Nano Lett. **3,** 193-198 (2003).
12. A. Javey, J. Guo, M. Paulsson, Q. Wang, D. Mann, M. Lundstrom, and H. Dai, Phys. Rev. Lett. **92,** 106804 (2004).
13. X. Cui, M. Freitag, R. Martel, L. Brus, and P. Avouris, Nano Lett. **3,** 783-7 (2003).
14. J. A. Misewich, R. Martel, P. Avouris, J. C. Tsang, S. Heinze, and J. Tersoff, Science **300,** 783-6 (2003).
15. *CRC Handbook of Chemistry and Physics* (CRC Press, Boca Raton, FL, 1996).
16. J. Appenzeller, M. Radosavljevic, J. Knoch, and P. Avouris, Phys. Rev. Lett. **92,** 483011-483014 (2004).
17. J. Nygard, D. H. Cobden, M. Bockrath, P. L. McEuen, and P. E. Lindelof, Appl Phys A **69,** 297-304 (1999).
18. B. Shan and K. Cho, Phys. Rev. B **70,** 233405-1-4 (2004).
19. P. Nikolaev, M. J. Bronikowski, R. K. Bradley, F. Rohmund, D. T. Colbert, K. A. Smith, and R. E. Smalley, Chem. Phys. Lett. **313,** 91-97 (1999).

49

**FIGURE CAPTIONS**

**FIG. 1.** Electrical properties of Rh contacted SWNT FETs: (a) Current vs. gate-voltage $I_{ds}$ vs. $V_{gs}$ data recorded at a bias of $V_{ds}$ =-600 mV (b) $I_{ds}$ vs. $V_{ds}$ curves at various gate-voltages labeled. (a) and (b) are for the same $d$~2.1 nm tube. (c) $I_{ds}$ vs. $V_{gs}$ ($V_{ds}$ = -500 mV) and (d) $I_{ds}$ vs. $V_{ds}$ curves for a $d$~1.5 nm tube. Insets in (a) and (c): AFM images of the devices, scale bars =200 nm.

**FIG. 2.** Diameter dependent electrical properties of semiconducting nanotube devices. (a) $I_{on}$ vs. $d$ for various semiconducting tubes. $I_{on}$'s are recorded at $V_{ds}$ = -1 V and $|V_g - V_{th}|$ = 3 V from $I_{ds}$ vs $V_{ds}$ curves of each device. (b) A schematic band diagram showing the Schottky barrier height at a metal-tube contact. The SB height follows Eq. (1) in the text for Pd and Rh and SB width in on the order of oxide thickness $t_{ox}$ of the device.

**FIG. 3.** A $d$ ~ 1.2 nm metallic nanotube. (a) $I_{ds}$ vs. $V_{gs}$ at room temperature. Inset: AFM image of the device. Scale bar = 200 nm. (b) $G$ vs. $V_{ds}$ shows non-linearity. (c) $I_{ds}$ vs. $V_{gs}$ measured at various temperatures indicated. $V_{ds}$ = 1 mV. (d) Differential conductance $dI_{ds}/dV_{ds}$ vs. $V_{ds}$ and $V_{gs}$ at 2K (Color scales from 0 to $0.1e^2/h$ from white to black).

**FIG. 4.** Diameter dependent electrical properties of metallic nanotubes. (a) Room temperature low-bias linear conductance $G$ measured for various $d$ metallic tubes. (b) $I_{on}$ vs. $d$ plots for both metallic and semiconducting SWNT devices with Rh or Pd contact. $I_{on}$'s all taken at $|V_{ds}|$ = 1V.

Figure 1

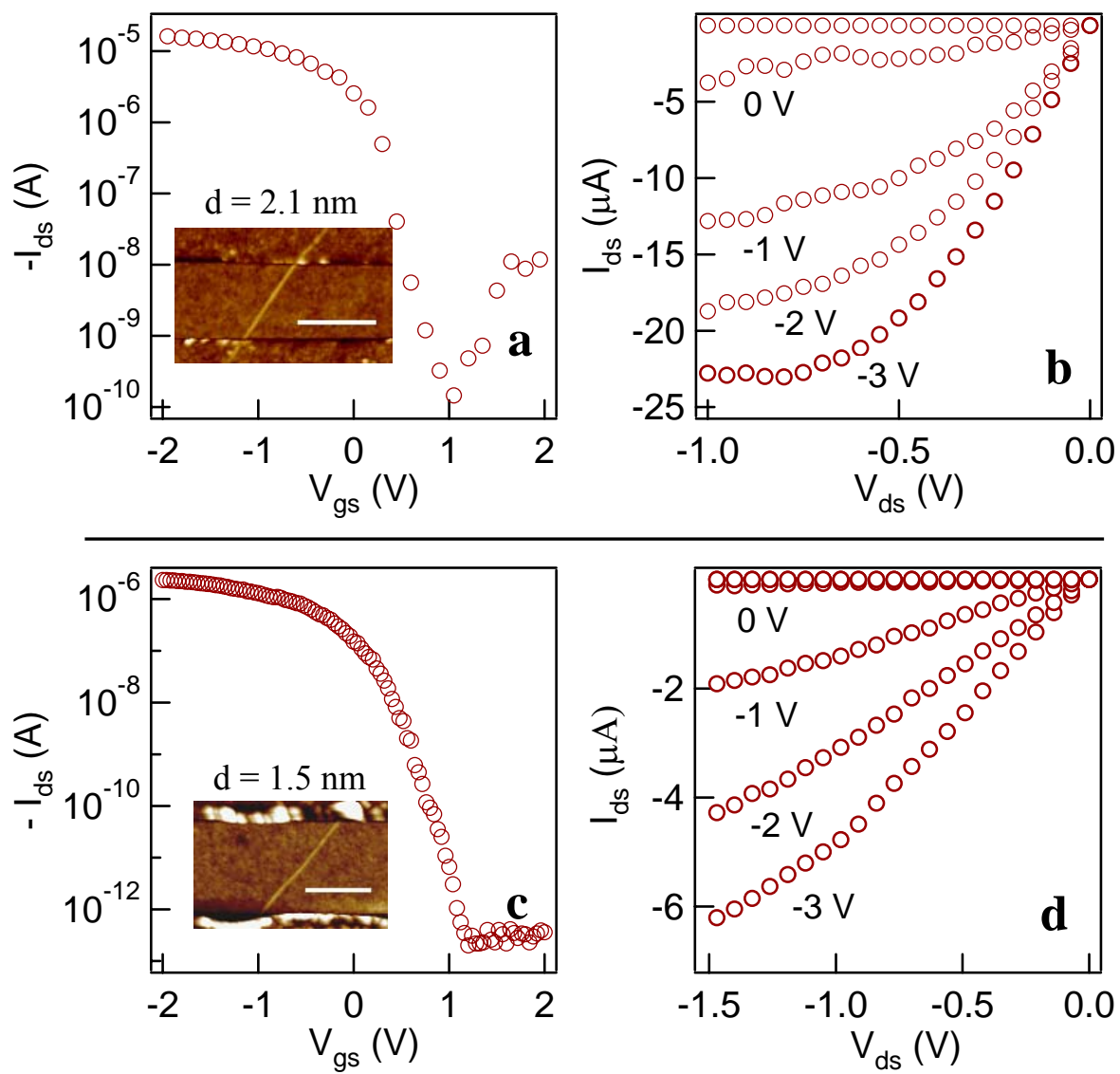

Figure 2

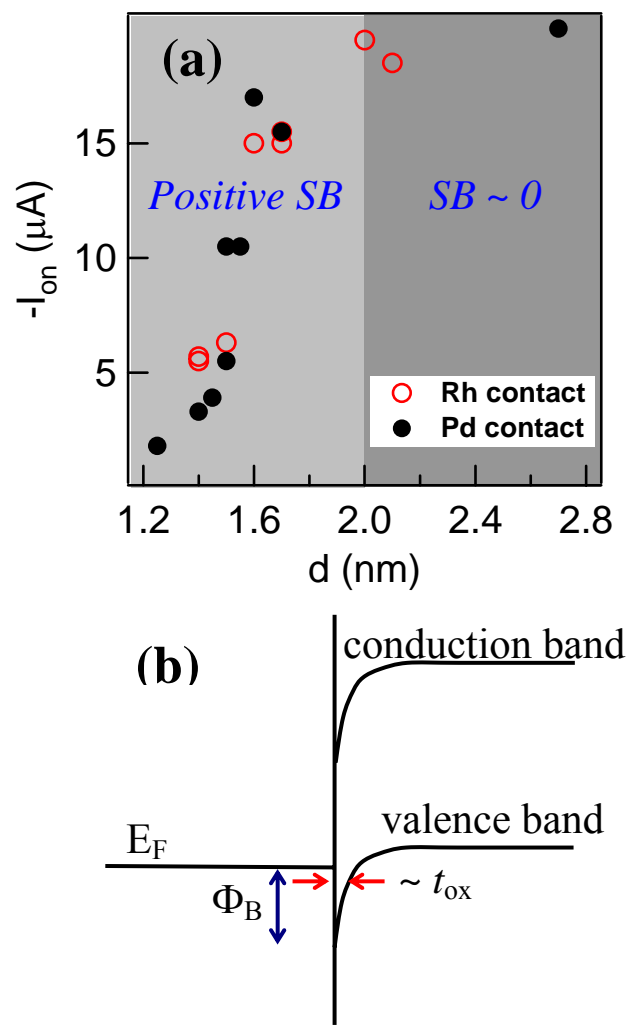

Figure 3

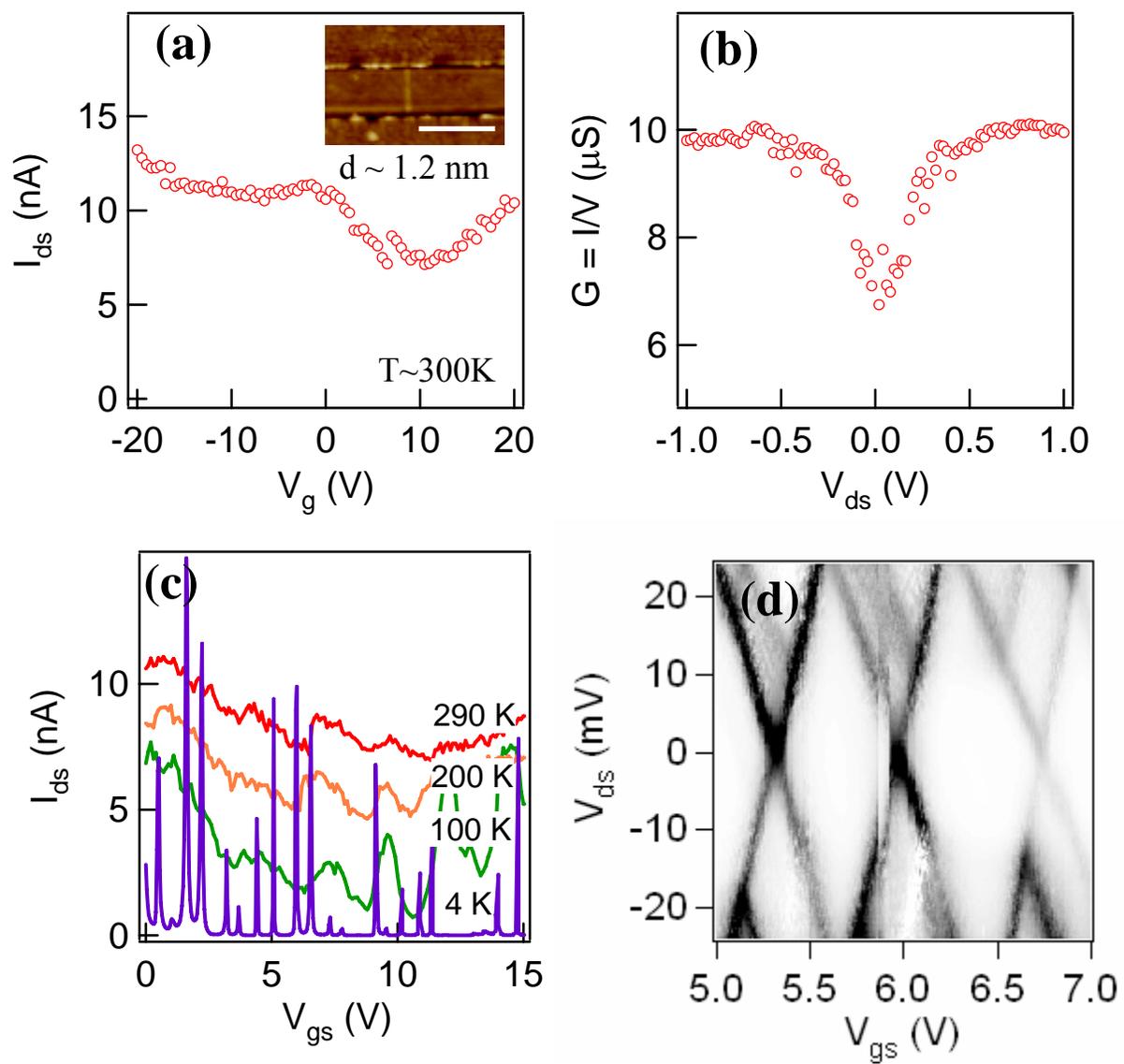

Figure 4

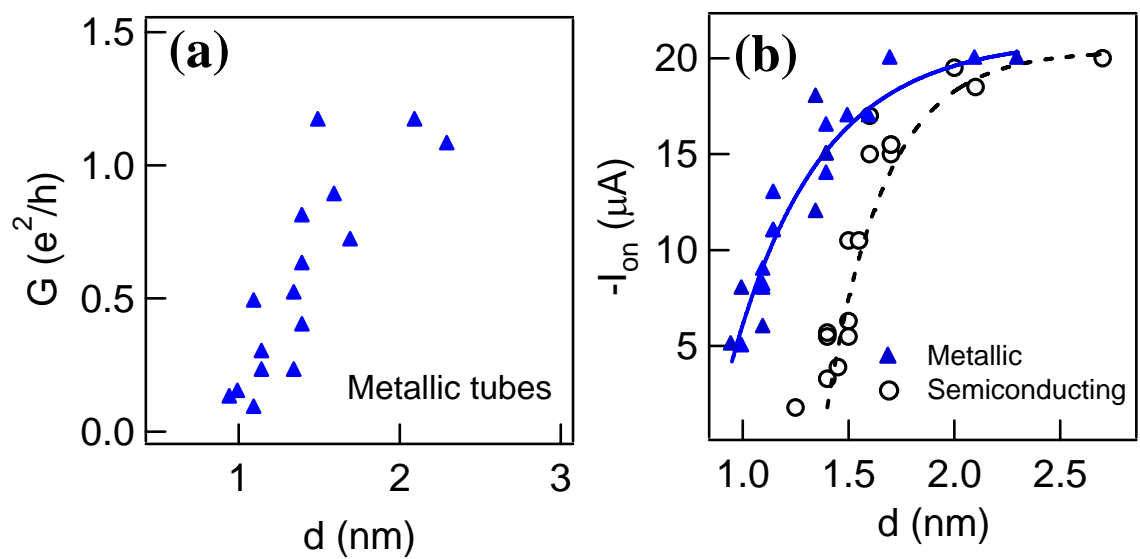